\documentclass[prx, twocolumn, superscriptaddress]{revtex4-1}
\usepackage{epsfig}
\usepackage{color}
\usepackage{mhchem}
\usepackage[makeroom]{cancel} 
\usepackage[normalem]{ulem} 
\usepackage{xcolor} 
\pagecolor{white}
\usepackage{amssymb,graphics,graphicx,amsfonts,amsmath,empheq}
\usepackage[breakable,most,many]{tcolorbox}
\usepackage{varwidth}
\usepackage{capt-of}
\usepackage[export]{adjustbox}

\usepackage{letltxmacro}
\LetLtxMacro{\originaleqref}{\eqref}
\renewcommand{\eqref}{Eq.~\originaleqref}

\begin{document}
\author{Tommaso Cossetto}
\affiliation{The Max Planck Institute of Molecular Cell Biology and Genetics, 01307 Dresden, Germany}
\affiliation{Gulbenkian Institute of Molecular Medicine, Oeiras, Portugal}
\affiliation{Current affiliation: Complex Systems and Statistical Mechanics, Department of Physics and Materials Science, University of Luxembourg}

\author{Jonathan Rodenfels}
\email[Corresponding author: ]{rodenfels@mpi-cbg.de}
\affiliation{Gulbenkian Institute of Molecular Medicine, Oeiras, Portugal}

\author{Pablo Sartori}
\email[Corresponding author: ]{pablo.sartori@gimm.pt}
\affiliation{Gulbenkian Institute of Molecular Medicine, Oeiras, Portugal}

\title{Charting dissipation across the microbial world}

\begin{abstract}
The energy dissipated by a living organism is commonly identified with heat generation. However, as cells exchange metabolites with their environment they also dissipate energy in the form of chemical entropy. How dissipation is distributed between exchanges of heat and chemical entropy is largely unexplored. Here, we analyze an extensive experimental database recently created \cite{Cossetto2024_thermo} to investigate how microbes partition dissipation between thermal and chemical entropy during growth. 
We find that aerobic respiration exchanges little chemical entropy and dissipation is primarily due to heat production, as commonly assumed.
However, we also find several types of anaerobic metabolism that produce as much chemical entropy as heat.  
Counterintuitively, instances of anaerobic metabolisms such as acetotrophic methanogenesis and sulfur respiration are endothermic.
We conclude that, because of their metabolic versatility, microbes are able to exploit all combinations of heat and chemical entropy exchanges that result in a net production of entropy.
\end{abstract}

\maketitle

Life is unavoidably linked to energy and its dissipation. In a cell, free energy is processed through metabolism to drive cellular functions, and is dissipated by production of entropy. Heat is often assumed to be the means by which cells dispense of the entropy produced by cellular activity. 
However, since cells exchange metabolites with their environment, the chemical entropy of these molecules also contributes to dissipation \cite{Roels}. 
In fact, such exchanges of chemical entropy might even prevail over heat production in certain types of metabolism \cite{heijnen1992search,von1999does,liu2001microbial}.

Given the versatility of microbial metabolism, unicellular organisms are ideal candidates to investigate the prevalence of different dissipation strategies, i.e. different partitioning of dissipation between heat and chemical entropy exchanges.
 In this report, we use the experimental database in \cite{Cossetto2024_thermo} to chart how different types of microbial metabolism dispense of dissipation during growth. We reach the conclusion that all possible strategies are exploited to grow in the microbial world.

\subsection*{Dissipation strategies of cellular growth}
\begin{figure}[]
	\includegraphics[]{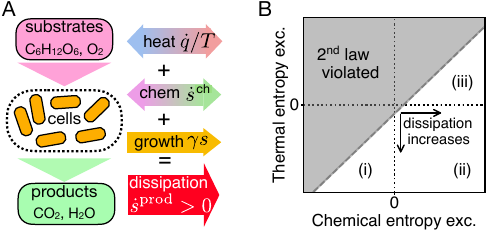}
	\caption{\label{fig:sketch}
		{\it Dissipation strategies of microbial growth.} {\bf A} Schematic of microbial cells growing by exchanging metabolites (left), and the entropy balance of steady growth, \eqref{eq:entrbal_sbg} (right). {\bf B} Space representing normalized exchanges of thermal entropy ($\dot q/T\gamma |s|$) on the vertical axis and of chemical entropy ($\dot s^{\rm ch}/\gamma |s|$) on the horizontal axis. The quadrants correspond to the possible dissipation strategies: (i) heat is produced and chemical entropy is decreased; (ii) heat and chemical entropy are produced; (iii) heat is absorbed and chemical entropy is produced. The gray area corresponds to violations of the second law, and the dashed line to $\dot s^{\rm prod} = 0$ (note the offset due to the r.h.s of \eqref{eq:entrbal_sbg_norm}).}
\end{figure}

Cells are bound by the second law of thermodynamics to dissipate free energy or, equivalently, to produce entropy at a positive rate,  $\dot{s}^{\rm prod}\ge0$. 
Consider a population of microbial cells that grows at a constant rate, $\gamma$, consuming substrates that are metabolized into products released to the environment, Fig.~\ref{fig:sketch}A. This process results in exchanges of heat at rate $\dot q$ ($<0$ when heat flows from the cells to the aqueous medium) and of chemical entropy at rate $\dot s^{\rm ch}$. The biomass of the population itself has an entropy content, which is determined by the cellular composition and, during balanced growth, is kept at a constant density, $s$. Here we consider entropies of formation and therefore $s<0$ \cite{Cossetto2024_thermo}. The second law can be written as a stationary entropy balance:
\begin{align}\label{eq:entrbal_sbg}
	\dot{s}^{\rm prod} = \gamma s -\dot{q}/T+\dot{s}^{\rm ch} \ge 0  \quad,
\end{align}
where $\dot{s}^{\rm prod}$, $s$, $\dot{q}$, and $\dot{s}^{\rm ch}$ are all defined per unit of biomass volume, and we assumed a constant temperature $T$ and fixed concentrations of chemicals in the environment.

\eqref{eq:entrbal_sbg} shows that to grow ($\gamma s$), cells exchange entropy with the environment in thermal and chemical form ($\dot{q}/T$ and $\dot{s}^{\rm ch}$). The contribution to dissipation of these two forms depends on the substrates and products of growth, i.e. on the type of metabolism. To compare different metabolic types in spite of their diverse growth rates, we rewrite \eqref{eq:entrbal_sbg} as
\begin{align}\label{eq:entrbal_sbg_norm}
	- \frac{\dot{q}}{T \gamma |s|} + \frac{\dot{s}^{\rm ch}}{\gamma |s|} \ge 1 \quad,
\end{align}
where the absolute value accounts for the negative sign of the entropy of biomass. We distinguish three dissipation strategies of microbial growth, Fig.~\ref{fig:sketch}B, according to the signs of the two terms in \eqref{eq:entrbal_sbg_norm}: 
(i) $\dot{q}<0$, $\dot{s}^{\rm ch}<0$:  growth produces heat but decreases chemical entropy; (ii) $\dot{q}<0$, $\dot{s}^{\rm ch}>0$: growth dissipates in the form of both heat and chemical entropy; (iii) $\dot{q}>0$, $\dot{s}^{\rm ch}>0$: cells absorb heat from the environment, but the release of chemical entropy is sufficient to fulfill the second law.

\begin{figure*}[t]
\includegraphics[]{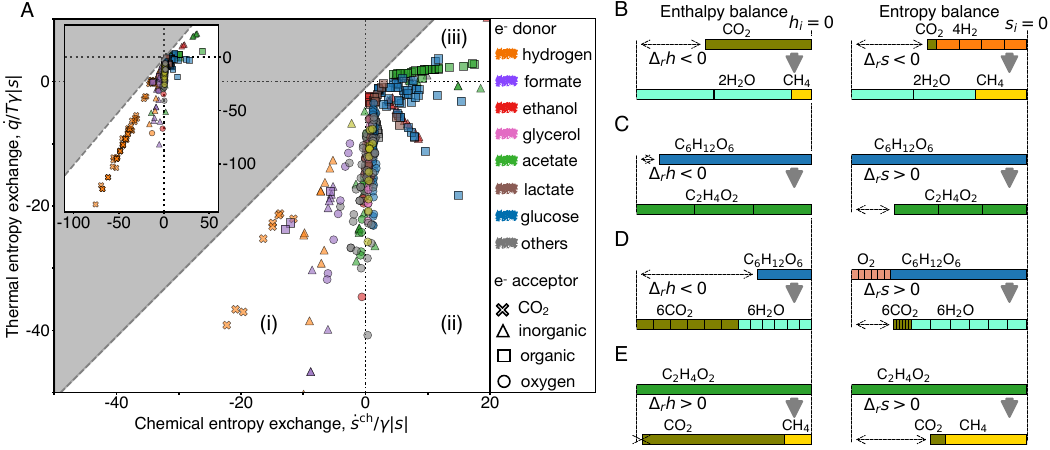}
\caption{\label{fig:datastrat}
{\it Experimental chart of dissipation strategies.} {\bf A} Scatter plot of the two terms in \eqref{eq:entrbal_sbg_norm} 
for diverse types of microbial growth taken from the database in \cite{Cossetto2024_thermo} ({\bf inset} for full range of all data). 
Each of the quadrants labeled (i), (ii) and (iii) corresponds to a dissipation strategy, all populated by a variety of metabolic types.
Strategy (i) is mainly populated by metabolic types using hydrogen and formate as electron donor (orange and purple symbols). Strategy (ii) is populated by different types of anaerobic fermentation and respiration (squares and triangles of various colors). Strategy (iii) is largely populated by acetotrophic methanogenesis (green squares). In addition, the transition region between strategies (i) and (ii) is occupied by aerobic types. {\bf B-E} Enthalpy and entropy balances ($\Delta_{\rm r} h$ left and $\Delta_{\rm r} s$ right respectively) of chemical reactions corresponding to representative metabolic types, with reactants on top and products at the bottom. 
The size of the colored bars is proportional to the enthalpy/entropy of formation (in standard conditions) of the molecules, which is negative for all molecules of all panels ($h_{{\rm H}_2}=h_{{\rm O}_2}=0$ by definition). The two dashed vertical lines crossing the panels mark the reference for the balance ($h_i=0$ and $s_i=0$), with all bars lying to their left.
 (B) Carbon dioxide methanation, \eqref{eq:methanation}, is exothermic ($\Delta_{\rm r} h <0$) but decreases chemical entropy ($\Delta_{\rm r} s<0$). 
 (C) Glucose breakdown to acetate, \eqref{eq:ferm}, produces little heat and chemical entropy in the absence of carbon dioxide. 
 (D) Glucose combustion, \eqref{eq:glucose}, is exothermic and produces little chemical entropy. 
 (E) Acetate methanation, \eqref{eq:met2}, is slightly endothermic but produces chemical entropy. }
\end{figure*}

Which of these dissipation strategies are found in nature? To answer this question, we studied a large dataset of metabolic dissipation of growing microbes spanning bacterial, archeal, and eukaryotic species \cite{Cossetto2024_thermo}. Following the approach detailed in  {\it Materials and Methods}, we computed the different terms in \eqref{eq:entrbal_sbg_norm}.  Figure~\ref{fig:datastrat} shows how dissipation decomposes into  heat and chemical entropy exchanges in each of these experiments. In the following, we illustrate the main metabolic types exploiting each strategy. Our main finding is that microbial metabolisms use all physically available dissipation strategies.

\subsection*{Strategy (i): heat produced, chemical entropy decreased}  
Metabolic types growing on one-carbon molecules, such as carbon dioxide and formate, exploit this strategy. 
Among these, autotrophic methanogenesis is utilized by archeal species that thrive at high temperatures (orange crosses in Fig.~\ref{fig:datastrat}A). In this type of metabolism, carbon dioxide is reduced either partially to produce biomass, or fully to methane. 
Due to the low biomass yield of this metabolism \cite{Cossetto2024_thermo}, the energy balance of the growth process is well approximated by the chemical reaction of carbon dioxide methanation:
\begin{align}\label{eq:methanation}
4 {\rm H}_2 + {\rm C}{\rm O}_2 \to {\rm C} {\rm H}_4 + 2 {\rm H}_2{\rm O}\quad.
\end{align}
In ``standard biological" conditions, this reaction is spontaneous because overall free energy is dissipated, i.e. entropy is produced. This results from how the entropic and enthalpic content of the molecules, making up the free energy, changes from reactants to products. As illustrated in Fig.~\ref{fig:datastrat}B, high entropy carbon dioxide is transformed into low entropy methane, with hydrogen consumption and water production further decreasing the chemical entropy. In contrast, the enthalpy balance is strongly negative because of the low enthalpy of water, despite methane having a higher enthalpy than carbon dioxide. Overall,  autotrophic methanogenesis compensates the decrease in chemical entropy by producing large amounts of heat.

\subsection*{Strategy (ii): heat and chemical entropy produced}  
This strategy is populated by anaerobic respiratory and fermentative metabolic types that dissipate a comparable amount of heat and chemical entropy. 

For example, glucose can be fermented into a variety of organic molecules (acetate, ethanol, butyrate, etc.) having an analogous entropy content. For reference, Fig.~\ref{fig:datastrat}C shows the breakdown of glucose to acetate,
\begin{align}\label{eq:ferm}
	{\rm C}_6{\rm H}_{12}{\rm O}_6 \to 3 {\rm C}_2 {\rm H}_4 {\rm O}_2 \quad .
\end{align}
However, the growth process also oxidizes glucose to carbon dioxide, which has a particularly high entropy (gram per gram), thus increasing the production of chemical entropy.

\subsection*{Transition region (i-ii): heat produced, chemical entropy negligible} 
We find a band of data in the transition region between strategies (i) and (ii), constituted by many types of aerobic respiration (circles of different colors in Fig.~\ref{fig:datastrat}A). Here the dissipation is dominated by heat production, and the exchange of chemical entropy is very small, either positive or negative. 

The energy balance of aerobic respiration differs from the reference chemical reaction, due to the high biomass yield \cite{Cossetto2024_thermo}.  
However, to gain a qualitative intuition, we consider the reaction of glucose combustion
\begin{align}\label{eq:glucose}
	{\rm C}_6{\rm H}_{12}{\rm O}_6 + 6 {\rm O}_2 \to 6 {\rm C} {\rm O}_2 + 6 {\rm H}_2{\rm O} \quad.
\end{align}
Fig.~\ref{fig:datastrat}D shows how the enthalpy balance is dominated by water formation from oxygen, with little chemical entropy either produced or consumed, similar to the overall growth process (Fig.~\ref{fig:datastrat}A). 

\subsection*{Strategy (iii): heat absorbed, chemical entropy increased}  
Endothermic growth operates in the regime opposite to strategy (i). Acetotrophic methanogenesis, used by several archeal species, has been previously shown to be endothermic \cite{liu2001microbial} in agreement with our results. Cells with this type of metabolism (green squares in Fig.~\ref{fig:datastrat}A) transform acetate into biomass, carbon dioxide, and methane, the two latter molecules being the most oxidized and reduced forms of carbon. The chemical reaction associated to this process, acetate disproportionation
\begin{align}\label{eq:met2}
{\rm C}_2 {\rm H}_4 {\rm O}_2 \to {\rm C} {\rm O}_2 + {\rm C} {\rm H}_4\quad,
\end{align}
absorbs $\sim10~{\rm kJ}$ of heat per mole of acetate. This is due to the high enthalpy of methane compared (gram per gram) to acetate and carbon dioxide, Fig.~\ref{fig:datastrat}E.  As with other methanogenic types, due to the low biomass yield \cite{Cossetto2024_thermo}, this is a good estimate for the heat absorbed during growth. 
	
Furthermore, we predict that sulfur respiration and certain types of sulfate respiration are also endothermic. In particular, \emph{Desulfuromonas acetoxidans}, a bacteria unable to ferment organic substrates, grows with solid sulfur as electron acceptor in two types of metabolism that appear to be strongly endothermic (Fig.~\ref{fig:datastrat}A inset, red and green triangles).

\subsection*{Discussion} Cellular dissipation cannot be reduced to heat production, because exchanges of chemical entropy can be equally or more relevant than heat. 
Despite having been known for decades \cite{Roels,heijnen1992search,von1999does}, this simple yet striking fact remains surprisingly overlooked.
By analyzing a large body of microbial growth data, we showed that microbes exploit all thermodynamically possible strategies to dissipate during growth. 
 Which dissipation strategies are exploited beyond laboratory conditions and outside the microbial world remains unknown. 

\subsection*{Methods}
We consider a population of cells dividing asynchronously in aqueous media. The media contains metabolites, labeled with index $i$, which are exchanged with the cells during growth at rates $f_i$ (per unit volume of biomass, positive for nutrients and negative for products). In a given experiment, these chemicals are the substrates and products of a particular metabolic type, for example in glucose respiration $i \in\{{\rm C}_6{\rm H}_{12}{\rm C}_6, {\rm O}_2, {\rm NH}_3, {\rm CO}_2, {\rm H}_2{\rm O}\}$, where the first three are substrates and the last two are products. We assume that the media acts as a thermal and chemical reservoir. 

During steady balanced growth, the total mass and volume of the population increase exponentially at the same constant rate $\gamma$. Under these conditions, the concentration of cellular constituents is constant, as well as the density rates of chemical exchanges $f_i$. Such conditions, met in batch cultures during exponential growth and in continuous (chemostat) cultures at steady state, directly result in the entropy balance of \eqref{eq:entrbal_sbg}.

To compute the entropy production rate, $\dot{s}^{\rm prod}$, we need explicit expressions for the rate of chemical entropy exchange, $\dot{s}^{\rm ch}$, and of heat exchange, $\dot{q}$. The former is given by
$
	\dot{s}^{\rm ch} = -\sum_i s_i f_i
	$,
with $s_i$ the entropy of formation of molecule $i$. The heat is similarly determined by the enthalpy balance,
$
	\dot{q} = \gamma h-\sum_i h_i f_i  
	$,
with $h_i$ the enthalpy of formation of the molecule $i$, and $h$ the enthalpy density of biomass. The enthalpy balance is a formulation of the first law of thermodynamics for microbial steady growth. 

\subsection*{Dataset} 
The dataset comprises 504 entries and is described in detail in \cite{Cossetto2024_thermo}. Briefly, each entry includes the yields (ratio of fluxes) of biomass, substrates, and products, extracted from published records of microbial growth experiments. In each entry, missing yields of nutrients and products were inferred using element conservation, see SI. The entropy balance in \eqref{eq:entrbal_sbg} was computed using 
enthalpies $h_i$ and entropies $s_i$ of formation of the molecules in ``standard biological'' conditions. The entropy and enthalpy of formation of biomass were computed using empirical formulas that relate $s$ and $h$ to element composition of biomass, see \cite{Battley1999, cordier1987relationship}, respectively.

\subsection*{Acknowledgments} This project has received funding from the ERC under the Horizon 2020 research and innovation program (Grant agreement No. 949811 — EnBioSys) to J.R.

\bibliographystyle{unsrtnat}
\bibliography{dissiallobiblio}

\vfill\eject\clearpage

	\section*{Supplementary Information}
	\subsection*{Extended Methods}
	Cells grow taking substrates from their environment and releasing metabolic byproducts. We label these molecules with the index $i$. The set of substrates and products, $\{{i}\}$, defines the metabolic type enabling growth.  Since mass is conserved, the overall biomass of the cellular population, $M$, changes over time according to the balance of mass exchanges with the media
	\begin{align}\label{eq:mass_balance}
		\frac{dM}{dt} &=  \sum_i	m_i F_i \quad ,
	\end{align}
	where $F_i$ is the flux of chemical $i$ exchanged between the cells and the media, and $m_i$ the mass of the corresponding molecules. Fluxes of substrates are positive, from the environment to the cell, and fluxes of products are negative, from the cell to the environment. 
	
	In the regime of balanced growth, the number of every cellular constituent (metabolites, proteins, complexes, etc.) increases at the same rate, $\gamma$, as well as the total biomass, giving rise to exponential growth
	$
	dM/dt = \gamma M 
	$.
	An analogous equation holds for the overall biovolume, $V$. As a consequence, the mass balance \eqref{eq:mass_balance} can be written in terms of time-independent quantities,
	\begin{align}\label{eq:mass_balance_density}
		\sum_i	m_i f_i &= \gamma \rho \quad ,
	\end{align}
	where $f_i = F_i / V$ and $\rho=M/V$ is the biomass density. We refer to the right hand side of this equation as the biomass flux. 
	
	In biochemical reactions, not only the total mass of the reactants is conserved but also the atoms, or elements, composing it. This allows to decompose \eqref{eq:mass_balance_density} into $K$ equations, one for each element 
	\begin{align}\label{eq:element_balance_density_SI}
		\sum_i	e^k_i f_i &= \gamma b^k \quad ,
	\end{align}
	where $e^k_i$ specifies the composition of molecule $i$ in element $k$ ($k =$ C, H, O, \dots), and $b_k$ is the concentration of element $k$ in the biomass.
		
	Because of the constraints imposed by \eqref{eq:element_balance_density_SI}, only $I+1-K$ of the fluxes of chemicals and biomass are independent, where $I$ is the total number of metabolites $i$ in the metabolic type. For each metabolic type, we use the biomass and other $I-K$ fluxes, labeled by $j$, as a base of independent fluxes to express the remaining fluxes
	\begin{align}\label{eq:dependent_fluxes}
		f_d = \sum_k  (e^{-1})^k_d \left( \gamma b^k - \sum_j e^k_j f_j \right) \quad ,
	\end{align}
	with $\sum_k (e^{-1})^k_d (e)^k_{d'} = \delta_{dd'}$, and $d, d'$ labeling dependent fluxes. 
	In the experimental dataset, the are always at least two independent fluxes, and roughly half of the data points belong to highly constrained metabolic types, i.e. with only two independent fluxes. Without loss of generality, we consider as independent the biomass and electron donor fluxes, which ratio is the biomass yield $y = \gamma b^C / |f_{\rm ed}|$. Note that the biomass yield is the quantity more often reported in the original sources of the data.
	
\subsection*{Example}
As an example, consider the aerobic growth of \emph{Escherichia coli} on glucose. The molecules exchanged with the media are $i =$ C$_6$H$_{12}$O$_6$, O$_2$, NH$_3$, CO$_2$, H$_2$O. 
There are $K=4$ elements (C, H, O, N), $I=5$ fluxes of metabolites plus the flux of biomass. Therefore, there are two independent fluxes. 

The experimental data typically available consists in the biomass yield on glucose, $y = \gamma b^C / |f_{ \rm{C}_6\rm{H}_{12}\rm{O}_6}|$ in units of carbon-moles per moles, and the elemental composition of biomass normalized to carbon content, i.e. $b^H/b^C, b^O/b^C, b^N/b^C$.

 We set biomass and glucose to be independent fluxes, because they are measured. All other fluxes are dependent, $d =$ O$_2$, NH$_3$, CO$_2$, H$_2$O, and can be computed in the form of yields, $y_d = f_d / |f_{\rm{C}_6\rm{H}_{12}\rm{O}_6}|$, dividing both sides of \eqref{eq:dependent_fluxes} by $f_{\rm{C}_6\rm{H}_{12}\rm{O}_6}$.

From its elemental composition, we can compute the enthalpy and entropy of formation of biomass, conveniently expressed in units of energy per carbon moles of biomass.  
Divided by $f_{\rm{C}_6\rm{H}_{12}\rm{O}_6}$, the second law \eqref{eq:entrbal_sbg} is expressed in terms of yields
\begin{align}\label{eq:entrbal_yields}
	\frac{\dot{s}^{\rm prod}}{|f_{\rm{C}_6\rm{H}_{12}\rm{O}_6}|} = \sum_i y_i (\frac{h_i}{T} - s_i) + y (s - \frac{h}{T}) \ge 0 \quad.
\end{align}
Finally, using the tabulated entropies and enthalpies of formation for the five metabolites, we can divide \eqref{eq:entrbal_yields} by $y \cdot s$ to obtain \eqref{eq:entrbal_sbg_norm} as plotted in Fig.~\ref{fig:datastrat}. 

\subsection*{Extended Dataset}

The curation of the dataset is explained in detail in \cite{Cossetto2024_thermo}. 
The entropy of formation $s_i$ of molecule $i$ at temperature $T$ is computed from its enthalpy of formation $h_i$ and free energy of formation $g_i$ according to the relation
\begin{align}
	s_i = \frac{1}{T} (h_i - g_i) \quad. 
\end{align}
The free energies of formation $g_i$, and therefore the entropies $s_i$, assume a concentration of $1$ mM as for ``standard biological" conditions.

In Fig.~\ref{fig:datastrat}D-E,  
$h_{ {\rm C}_2 {\rm H}_4{\rm O}_2, {\rm \, liq} } = -483.52$ (kJ/mol) is from the NIST,
 and $g_{{\rm C}_2 {\rm H}_4{\rm O}_2, {\rm \, aq}} = -396.48$ 
(kJ/mol) from Jan P. Amend, Everett L. Shock, Energetics of overall metabolic reactions of thermophilic and hyperthermophilic Archaea and Bacteria, FEMS Microbiology Reviews, Volume 25, Issue 2, April 2001, Pages 175–243.

\subsection*{Endothermic data}
In the main text, we discussed the fermentation of acetate to methane, the respiration of solid sulfur to oxidize acetate or ethanol, and mentioned the respiration of sulfate to oxidize acetate, propionate, or butyrate. 
Not mentioned in the main text, there is one data point of fumarate reduction with acetate oxidation that gives a slightly endothermic balance.
Finally, four data points of glucose fermentation by \emph{Clostridium butyricum} give an endothermic enthalpy balance. These data points involve many fermentation products (acetate, hydrogen, butyrate, ethanol, etc.), and therefore the reconstruction of the yields through element balance is difficult. These yields are critical to determine the value of the enthalpy balance, and since more points belonging to the closely related metabolic type are exothermic, we question the reliability of this data. 
	
\end{document}